\documentclass[aps,prd,reprint,superscriptaddress]{revtex4-1}

\usepackage[utf8]{inputenc} 
\usepackage{graphicx,color,overpic,mathtools}
\usepackage{amsthm,amsmath,amssymb,hyperref}
\usepackage{braket,bm,bbm,setspace}
\PassOptionsToPackage{normalem}{ulem}
\usepackage{ulem}

\newcommand{\op}{\hat}
\newcommand{\ft}{\tilde}
\newcommand{\ii}{\mathrm{i}}
\renewcommand{\d}{\mathrm{d}}

\begin{document}

\title{Thermalization of particle detectors: The Unruh effect and its reverse}

\author{Luis J. Garay}
\affiliation{Departamento de F\'isica Te\'orica II, Universidad Complutense 
de Madrid, 28040 Madrid, Spain}
\affiliation{Instituto de Estructura de la Materia (IEM-CSIC), Serrano 121, 28006 Madrid, Spain}

\author{Eduardo Mart\'in-Mart\'inez}
\affiliation{Institute for Quantum Computing, University of Waterloo, Waterloo, Ontario, N2L 3G1, Canada}
\affiliation{Department of Applied Mathematics, University of Waterloo, Waterloo, Ontario, N2L 3G1, Canada}
\affiliation{Perimeter Institute for Theoretical Physics, 31 Caroline St N, Waterloo, Ontario, N2L 2Y5, Canada}

\author{Jos\'e de Ram\'on }
\affiliation{Departamento de F\'isica Te\'orica II, Universidad Complutense 
de Madrid, 28040 Madrid, Spain}

\date{July 15, 2016}

\begin{abstract} 
We study the Anti-Unruh effect in general stationary scenarios. We find that, for accelerated trajectories, a particle detector coupled to a KMS state of a quantum field can cool down (click less often) as the KMS temperature increases. Remarkably, this is so even when the detector is switched on adiabatically for infinitely long times.
 We also show that the Anti-Unruh effect is characteristic of accelerated detectors, and cannot appear for inertially moving detectors (e.g., in a thermal bath).
 \end{abstract}

\maketitle

\section{Introduction}

It has been known since the 70s that uniformly accelerating particle detectors in flat spacetime, coupled to the vacuum state of a scalar field, would detect a thermal bath of particles \cite{Unruh1}. This phenomenon has become known as the Unruh effect. Despite the lack of direct experimental confirmation, it is regarded as one of the cornerstones of our understanding of quantum field theory.  The temperature  of this thermal bath is proportional to the magnitude   of the proper acceleration of the detector. In fact, accelerated detectors carefully switched on and coupled for long times to the Minkowskian vacuum thermalize \cite{fewsterjorma,Wilson}.  

However, recent work has shown that accelerated detectors in 1+1D spacetimes, in the presence of an infrared (IR) cutoff, and switched on for short times (below the detector's Heisenberg time) can actually {\it become cooler} instead of {\it warm up} in the sense that their clicking rate decreases as the Unruh temperature increases \cite{Anti-Unruh}.

To support this statement the authors of \cite{Anti-Unruh} use the Excitation-to-Deexcitation Ratio (EDR) to define a temperature estimator $  T_\textsc{edr}=-\Omega[\log(P_+/P_-)]^{-1},$  (where $P_\pm$ is the excitation/deexcitation probability and $\Omega$ is the detector's energy gap).   They   found that $T_\textsc{edr}$ was effectively  independent of the detector's energy gap when the Anti-Unruh effect is present, a behaviour usually associated with stationarity. The results of \cite{Anti-Unruh} raise several very interesting questions. In particular, is this unintuitive `cooling' effect just transient behaviour? Can it happen under true stationarity conditions? How general is this effect and what is its cause? 

In this work we analyze the ingredients that can  lead to these Anti-Unruh phenomena, study their relationship with the Kubo-Martin-Schwinger (KMS) condition \cite{Kubo,MartinS,fewsterjorma}, and single out general conditions under which it will or will not take place.

With this aim we will first study under what conditions the EDR temperature estimator $T_{\textsc{edr}}$  is   independent of detector's energy gap. We will show that this is actually, to a very good approximation, the case in a variety of  scenarios for short and long interaction times.

Furthermore, we will characterize the Anti-Unruh effect for general KMS states (as rigorously defined in \cite{fewsterjorma}) in terms of two different behaviours. As a first variety of this effect, we will show that under KMS, a detector may indeed click less often as the KMS temperature increases (\textit{weak} Anti-Unruh).  
More strikingly, as a second variety it may also happen that the effective EDR temperature is almost independent of the gap frequency and at the same time decrease as the KMS temperature increases (\textit{strong} Anti-Unruh).  

These effects appeared in \cite{Anti-Unruh} for  accelerated detectors coupled to a massless scalar field in two scenarios: \textit{a)} Under a hard-IR momentum cutoff in free 1+1D flat spacetime, and \textit{b)} for a detector in a  periodic cavity from which the zero mode is removed. 
In those cases the Wightman function is not stationary, thus, strictly speaking, the Wightman function is not KMS. 
However, we will show that the results in \cite{Anti-Unruh} are not due to this fact since the two modalities of Anti-Unruh phenomena mentioned above are also present in scenarios where the KMS condition is fully satisfied.

As an example, we analyze in detail the case of a uniformly accelerated detector coupled to the vacuum of a massive scalar field in 1+1D. We will see that, for small interaction times, the response function  decreases as the acceleration  increases. Moreover, we will also show the existence of regimes where the effective EDR temperature decreases as the  KMS temperature increases. The non-monotonicity of $T_\textsc{edr}$ as a function of the KMS temperature disappears for long interaction times or large temperatures. Surprisingly,  the observation that the transition probability can decrease as the detector's acceleration increases with the detector's acceleration survives even in the limit of infinitely long times under KMS. Therefore this modality of  Anti-Unruh effect cannot be associated with transient behaviour in any way.  

Interestingly, we will show that under the KMS condition and when the trajectory of the detector does not depend on the KMS temperature (i.e., inertial detectors interacting with a thermal state of a scalar field), the Anti-Unruh effect cannot  appear at all.  We will therefore show that the perception of the Anti-Unruh effect is linked to accelerated observers and it is not present for inertial observers coupled to generic thermal baths.

In Sec.~\ref{piposec:udw-kms} we first summarize the main concepts and tools concerning Unruh-DeWitt detectors interacting with scalar fields and the KMS condition that characterize equilibrium field states. Then we introduce the $T_{\textsc{edr}}$ temperature estimator and analyze its behaviour for long interaction times. Sec. \ref{piposec:au-kms} is devoted to the Anti-Unruh effect and its relation with the KMS condition. We first define the two modalities of weak and strong Anti-Unruh behaviours, and present necessary conditions for their appearance. Then we present, discuss, and compare different general situations with and without Anti-Unruh effect. We summarize and conclude in Sec. \ref{piposec:conc}

\section{Unruh-DeWitt detectors and KMS condition}
\label{piposec:udw-kms}

\subsection{Thermality and the KMS condition} 

 Thermal states in quantum statistical mechanics are described by Gibbs’ distribution. This distribution is well defined for systems with finite (or countably infinite) degrees of freedom. However, for systems of continuous variables, (e.g., quantum fields) the Gibbs distribution is not well defined \cite{Strocchi2008}. In those cases, it is still possible to approach the problem considering large---but finite---systems and then taking the thermodynamic limit.

The Kubo-Martin-Schwinger (KMS) condition \cite{Kubo,MartinS} was introduced in quantum statistical mechanics as a general abstract definition of equilibrium states that, for continuous variables systems, avoids the problems of having to see them as limiting cases of systems with finitely many degrees of freedom. As such, it has provided a more natural way to define equilibrium states in quantum field theory.

Let us consider the evolution of a free scalar quantum field $\op\phi(\mathsf{x})$  along a timelike vector $\partial_\tau$. If $\mathsf{x(\tau)}$ is the curve generated by $\partial_\tau$ for some initial condition $\mathsf{x}(0)$, the field at time $\tau$ can  be written as the pullback $\op\phi\big(\mathsf{x(\tau)}\big)$  of the field along the trajectory $\mathsf{x}(\tau)=\big(t(\tau),\bm x(\tau)\big)$.

The KMS condition for the field and the evolution parameter $\tau$ can be formulated as follows: 

We define the field's Wightman function 
$W  (\tau,\tau')$ in a given field state represented by $\op\rho$ as
\begin{equation}
    W  (\tau,\tau'):=\braket{\op\phi\big(\mathsf{x}(\tau)\big) \op\phi\big(\mathsf{x}(\tau')\big)}_{\op\rho}.
\end{equation}
where $\langle\cdot\rangle_{\op\rho}$ represents the expectation value on the state $\op\rho$. Then, this field state $\op\rho $ is said to  satisfy the KMS condition in the time parameter $\tau$ with inverse KMS temperature $\beta$ if and only if \cite{fewsterjorma}
\begin{equation}\label{eq:KMSperiod}
    W (\tau-\ii\beta,\tau' )=W (\tau',\tau).
\end{equation} 

The KMS condition connects the two-point correlations in the field with the stationarity of the field state. In fact, it can be shown  \cite{Kubo,MartinS}  that the KMS condition~\eqref{eq:KMSperiod} implies stationarity of the Wightman function (i.e., invariance under translations in $\tau$). In other words, \eqref{eq:KMSperiod} implies that the Wightman function depends only on the difference between its two arguments:
\begin{equation}\label{pipETA}
     W  (\tau,\tau')= W(\tau-\tau')= W(\Delta \tau).
\end{equation}
Notice that the converse is not true (that is, stationarity is necessary but not sufficient for KMS). It is easy to prove that Gibbs states are KMS states. For example, the thermodynamic limit of a Gibbs state of a scalar field defined in a finite volume satisfies the KMS condition even after taking the limit where the volume tends to infinity \cite{Haag,Takagi}. As such, the KMS condition  is a necessary condition for thermodynamic equilibrium. However, in general, it is not sufficient. Namely, equilibrium usually requires extra conditions, such as stability under perturbations. In general, equilibrium states are passive, and all KMS states are passive \cite{pusz1978}. The converse is only true under certain circumstances \cite{araki1974}. 

Let us now consider situations that lead to stationary Wightman functions. In these circumstances, the KMS condition can be formulated as an imaginary periodicity condition on $W(\Delta\tau)$ (see \cite{fewsterjorma} for a precise definition):
\begin{equation}
    \label{eq:KMSperiodest}
    W (\Delta\tau-\ii\beta )=W (-\Delta\tau).
\end{equation}  
If we Fourier transform this equation, we obtain the so-called detailed balance condition at inverse KMS temperature $\beta$:
\begin{equation}\label{detailpipo}
    \ft W(-\omega)=e^{\beta\omega}\ft W(\omega),
\end{equation}
where $\ft W$ is the Fourier transform of $W$
\begin{equation}
    \ft W(\omega)=\int_{-\infty}^\infty\!\!\!\d  \tau  e^{-i\omega \tau}W( \tau).
\end{equation}
 
Finally, the expectation value of the commutator 
 \begin{equation}
     C(\tau,\tau')=\braket{[\op\phi(x(\tau)),\op\phi(x(\tau'))]}_{\op \rho},
 \end{equation}
is the  imaginary part of the Wightman function. Hence, if $W$ is stationary $C$ will also be stationary  
 \begin{equation}
     C(\Delta\tau)= C(\tau-\tau')= 2\ii\,  \text{Im}\big(W(\Delta\tau)\big),
 \end{equation}
 Note that the (expectation value of the) commutator $C(\Delta\tau)$ is independent of the field state $\op \rho$, despite the fact that that the Wightman function is not.
 In terms of Fourier transforms, this equation becomes
 \begin{equation}
     \ft C(\omega)=\ft W(\omega)-\ft W(-\omega).
     \label{pipotador}
 \end{equation}
 This allows us to equivalently write the detailed balance condition as
 \begin{equation}
     \ft W(\omega,\beta)=-\ft C(\omega,\beta) \mathcal P(\omega,\beta),
     \label{eq:wcp}
 \end{equation}
 where $\mathcal P(\omega,\beta)$ is the Planckian distribution at inverse KMS temperature $\beta$,
 \begin{equation}\label{Piploanckiana}
     \mathcal P(\omega,\beta)=\frac1{e^{\beta\omega}-1}.
 \end{equation}
 Notice that in \eqref{eq:wcp} and from now on, when we write a Wightman function that satisfies the KMS condition we will write explicitly the dependence of $\ft W$ and $\ft C$ on the inverse KMS temperature $\beta$.

\subsection{Unruh-DeWitt detectors}

Particle detectors are localized and controllable quantum systems that are locally coupled in space and time to quantum fields. With the introduction of particle detectors we gain a tool to extract localized spatio-temporal information from the fields. Particle detector models in quantum field theory  were pioneered by Unruh and DeWitt \cite{Unruh1,DeWitts}, and can be found in the literature in several slightly different (but fundamentally similar) formats, e.g., a field in a box \cite{Unruh1}, a two-level system \cite{DeWitts} or a harmonic oscillator (see, e.g., \cite{BeiLok,Brown2012,Fuenetesevolution}). 

These models are commonly used in experimental setups in quantum optics \cite{scullybook} and in superconducting circuits \cite{Wallraff:2004aa}. For example, the orbital structure of an atom as a first quantized system can serve as such a detector for the second quantized electromagnetic field. Indeed, many common light-matter interaction models are fundamentally identical in nature to the Unruh-DeWitt (UDW) model \cite{scullybook}. In fact, the UDW model captures the features of the light-matter interaction in quantum optics for processes that do not involve exchange of orbital angular momentum \cite{Wavepackets,Alvaro,Pozas-Kerstjens:2016}. 

The UDW model consists of a real scalar field $\op\phi$ obeying the Klein-Gordon equation and a two-level system, whose ground $\ket g$ and excited $\ket e$ states are separated by an energy gap $\Omega$. Their coupling is described by an interaction Hamiltonian (in the interaction picture)
\begin{equation}
    H_I=\lambda\chi(\tau/\sigma)\mu(\tau) \op\phi(\mathsf{x}(\tau)),
\end{equation}
where   $\lambda$ is the coupling strength, $\chi (\tau/\sigma)$ is a  square integrable switching function satisfying $\|\chi\|_{_2}=1$, $\sigma$ is an interaction duration timescale,  $\mu(\tau)$ is the monopole moment of the detector, and $\mathsf{x}(\tau)$ is the spacetime trajectory of the detector parametrized by its proper time $\tau$. For a detector initially in the ground state, and the field in an initial arbitrary   state $\op\rho$, the excitation probability at leading order in the coupling strength~$\lambda$ turns out to be:
\begin{equation}
\label{eq:probability-udwplus}
P^+=\lambda^2 |\bra e\mu(0)\ket g|^2 \sigma\mathcal F(\Omega,\sigma),
\end{equation}
where
\begin{equation}
    \mathcal F(\Omega,\sigma)\!=\!\frac1\sigma\!\!\int_{-\infty}^\infty \!\!\!\!\!\!  \d \tau'\!\!\! \int_{-\infty}^\infty\!\!\!\!\!\! \d \tau \chi (\tau/\sigma)\chi(\tau'\!/\sigma){W}  ( \tau -\tau')e^{-\ii\Omega(\tau-\tau')}
\end{equation}
is the so-called response function of the detector. In terms of Fourier transforms, this expression can be written as 
\begin{equation}
    \mathcal F(\Omega,\sigma)=\frac{1}{2\pi} \int_{-\infty}^\infty\d\bar\omega |\ft \chi(\bar\omega)|^2 \ft W  (\Omega + \bar\omega/  \sigma  ),
    \label{eq:FOmega}
\end{equation}
where $\bar\omega$ is a dimensionless integration variable. Recall that $\tilde\chi$ is square integrable. However, for \eqref{eq:FOmega} to be well-defined for an arbitrary Wightman function we need to require additional conditions on how fast $\ft\chi(\omega)$ decays. For our purposes it will be sufficient to assume $\tilde\chi$ decays faster than a polynomial, which is a fairly mild assumption that covers the typical cases considered in the literature (e.g., $\ft \chi$ corresponding to Gaussian, Lorentzian or sinc switching functions  $\chi$). Note that the response function does not depend on the internal structure of the detector.

It is straightforward to check that the decay probability from the excited to the ground state is 
\begin{equation}
\label{eq:probability-udw}
P^-=\lambda^2 |\bra e\mu(0)\ket g|^2 \sigma\mathcal F(-\Omega,\sigma).
\end{equation}

Let us define another quantity of interest. We define the Excitation-to-Deexcitation Ratio (EDR) of the detector $\mathcal R(\Omega,\sigma)$ as the ratio between the excitation and decay probabilities:
\begin{equation}\label{pipoEDR}
    \mathcal R(\Omega,\sigma)=\frac{\mathcal F(\Omega,\sigma)}{\mathcal F(-\Omega,\sigma)}.
\end{equation}

Let us consider first the limit of arbitrarily long interaction time $\sigma$ and compute \eqref{eq:FOmega} in this limit. Recall that to take the limit  under the integral sign for each value of $\Omega$, and for an arbitrary Wightman function we need to impose conditions on the decay of $\ft\chi(\omega)$. Again, for our purposes it will be sufficient to assume that $\tilde\chi$ decays faster than a polynomial. The result of taking such limit is
\begin{equation}\label{hastalospipuevos}
    \mathcal F(\Omega,\sigma)\mathop{\longrightarrow}_{ \sigma\to\infty}\ft W(\Omega).
\end{equation}
In this limit, the EDR is 
\begin{equation}
    \mathcal R(\Omega,\sigma)\mathop{\longrightarrow}_{ \sigma\to\infty}\frac{\ft W(\Omega)}{\ft W(-\Omega)}.
\end{equation}
This means that we have pointwise convergence. However in can be shown  that the convergence is not uniform~\cite{fewsterjorma}.

If we further assume that the field state is KMS so that $\ft W$ satisfies the detailed balance condition \eqref{detailpipo}, we recover the  well-know expression for thermalized detectors
\begin{equation}\label{piporemark}
    \mathcal R(\Omega,\sigma)\mathop{\longrightarrow}_{ \sigma\to\infty}e^{-\beta\Omega}.
\end{equation}

\subsection{EDR and interaction time}

As the last part of this section let us analyze how the equilibrium EDR expression  \eqref{piporemark} is modified away from the infinitely long interaction time regime.  

In order to understand the thermalization process at finite times, we would like to study the response of particle detectors in the regime of long (but finite) interaction times $\sigma$. One possible approach is to try to find a representation of $\ft W$  as a power series on $\bar\omega/(\sigma\Omega)$.  The main problem is that, in general, $\ft W (\Omega+ \bar\omega/  \sigma ) $ is not an analytic function of $\bar\omega$,  (e.g. those satisfying the KMS condition). Let us call $\bar\omega_c$ the convergence radius of the power series expansion of $\ft W (\Omega+ \bar\omega/  \sigma ) $ centred at  $\bar\omega=0$. We can divide the integral \eqref{eq:FOmega} in two parts, one below and one above $\bar\omega_c$. Namely, we define
\begin{equation}
    \mathcal F_{\text{in}}(\Omega,\sigma)=\frac{1}{2\pi} \int_{-\bar\omega_c}^{\bar\omega_c}\d\bar\omega |\ft \chi(\bar\omega)|^2 \ft W(  \Omega+ \bar\omega/  \sigma ) ,
    \label{PipoFin}
\end{equation}
so that the response function \eqref{eq:FOmega} can be split in two parts as
\begin{equation}
 \mathcal F(\Omega,\sigma)=\mathcal F_{\text{in}}(\Omega,\sigma)+\mathcal F_{\text{out}}(\Omega,\sigma),
\end{equation}
where the integration range of $\mathcal F_{\text{out}}(\Omega,\sigma)$ is $|\bar\omega|>\bar\omega_c$.

Now, for $|\bar\omega|<\bar\omega_c$, $\ft W$ can be represented by the following power series 
\begin{equation}\label{31}
    \ft W\left(\Omega +\frac{\bar\omega}{ \sigma}\right) =\sum_{k=0}^{\infty} \frac{1}{k!} \frac{\d^{k}\ft W(\Omega)}{\d\Omega^{k}} \left(\frac{\bar\omega}{ {\sigma}}\right)^k.
\end{equation}

If we choose a `fat' switching function such that its Fourier transform $\ft\chi$ is compactly supported within the convergence radius $\bar\omega_c$ of the expansion \eqref{31}, we have that $\mathcal F_{\text{out}}(\Omega)=0$, so that $\mathcal F(\Omega)=\mathcal F_{\text{in}}(\Omega)$. Then, we can formally write
\begin{equation}\label{piposerie}
    \mathcal F(\Omega,\sigma)=  \sum_{k=0}^{\infty} \frac{\ft W^{(2k)}(\Omega)}{(2k)!} {\sigma}^{-2k}\int_{-\infty}^\infty \!\!\!\!\! \d\bar\omega\,\bar\omega^{2k}|\tilde{\chi}(\bar\omega)|^2.
\end{equation}
The remaining integral is 
just the $2k$-th moment of the Fourier transform of the switching function (squared).
Successive integrations by parts allow us to write it as the square of the norm   of the $k$-th derivative of the switching function $\|\chi^{(k)}\|_{_2}^2$, as can be checked straightforwardly. Then,
\begin{equation}\label{piposerie2}
   \mathcal F(\Omega,\sigma)= \ft W(\Omega)+ 
    \sum_{k=1}^{\infty} \|\chi^{(k)}\|_{_2}^2  \frac{\ft W^{(2k)}(\Omega)}{(2k)!}{\sigma}^{-2k} ,
\end{equation}
where we have singled out the infinite interaction time contribution.

Notice that we have implicitly assumed above that the integral of the series is the series of the integral. As we will discuss below, for the series of the integral to converge we need to impose extra conditions on the switching function. 

In particular, let us consider that the Fourier transform of the switching function $\ft \chi$ is compactly supported for $-A< \bar\omega<A$ with $A<\bar\omega_c$. Recall that $\chi$ is a square integrable switching function of $L^2$ norm unity. This means that $\ft\chi$ is bounded.  We can bound the $2k$-th moments of $\ft \chi(\bar\omega)$ as follows
\begin{equation}
 \frac{2A^{2k+1}\ft{X}_\text{inf}}{2k+1} \le   \int_{-A}^A \!\!\!\!\! \d\bar\omega\,\bar\omega^{2k}|\ft{\chi}(\bar\omega)|^2\le \frac{2A^{2k+1}\ft{X}_\text{sup}}{2k+1}, 
\end{equation}
where $\ft{X}_\text{inf}$ and $\ft{X}_\text{sup}$ are respectively the infimum and supremum of $|\ft\chi(\omega)|^2$ within its compact support.

If $\chi$ satisfies the assumptions above, we can see  that the series \eqref{piposerie} is convergent if $A\le1$. In other words, for switching functions whose Fourier transform is compactly supported for the interval $-1\le \bar\omega \le 1$, the series \eqref{piposerie} and \eqref{piposerie2} converge to the detector's response function provided that $\bar\omega_c\ge 1$.

The  set of switching functions that satisfy the conditions above is seriously restricted, although it still contains certain generality. Examples of these include general convolutions of smooth functions with a sinc function.

In the cases where \eqref{piposerie2} represents the detector's response, we can also write the EDR \eqref{pipoEDR} as a series expansion in order to study the thermalization behaviour of an Unruh-DeWitt detector.
The radius of convergence $\bar\omega_c$ of the power series of the Wightman function depends on its analytic structure and is specific to each particular case.

\subsubsection*{KMS massless field in 3+1 dimensions}

Let us consider the case of 3+1 dimensions and an inertial detector coupled to a massless scalar field in a KMS state with respect to the time parameter $\tau$ and inverse KMS temperature  $\beta$. For example, this could correspond to the case of an inertial detector in a thermal background at temperature $T$, where $\beta=1/T$ and $\tau$ is the proper time of the inertial detector. Equivalently \cite{Takagi}, this could correspond to the case of the field in vacuum and where $\tau$ is the proper time of an accelerated detector of proper acceleration $a$ and where $\beta=2\pi/a$. 

In these cases, the Fourier transform of the  Wightman function has the form \cite{Takagi}
\begin{equation}
    \ft W(\omega,\beta)=\frac{ \omega }{2\pi(e^{ \beta\omega}-1)}.
\end{equation}
The function $\ft W  (\Omega + \bar\omega/ \sigma ) $ presents singularities in the complex $\bar\omega$-plane, in particular when  $\beta ( \Omega+\bar\omega/\sigma)=2\pi n\ii$ with $n\in\mathbb Z$.
Hence the radius of convergence of \eqref{31} is $\bar\omega_c=\sigma \Omega  \sqrt{1+(2\pi/\beta\Omega)^2}$.

As discussed above, for the convergence of \eqref{piposerie2}, we need to demand that $\bar\omega_c>1$. This is guaranteed if \mbox{$\sigma\Omega>1$}, that is, this series expansion is guaranteed to converge to the detector's response function for times above the Heisenberg time of the detector $\Omega^{-1}$. The series is also valid for times below the Heisenberg time of the detector provided that $ \beta/\sigma<2\pi$.

In summary, for interaction times above the Heisenberg time of the detector the response function can be written as
\begin{equation}\label{PIPADA}
\mathcal F(\Omega,\sigma,\beta)\!=\! \ft W(\Omega,\beta)+
  \frac{\|\chi'\|_{_2}^2}{2} \partial_\Omega^2\ft W(\Omega,\beta)\sigma^{-2} +\mathcal{O}\!\left(\!\frac{1}{\Omega^4\sigma^4}\!\right)\!.
\end{equation}
From the EDR \eqref{pipoEDR}, we can now define an effective inverse  EDR temperature as
\begin{equation}\label{otrapipocosa}
    \beta_\textsc{edr}(\Omega,\sigma,\beta)=-\frac{\log{\big(\mathcal R(\Omega,\sigma,\beta)\big)}}{\Omega},
\end{equation}
which in the limit $\sigma\to\infty$ coincides exactly with the inverse KMS temperature $\beta$, as it can be seen from \eqref{piporemark}. Using the series expansion \eqref{PIPADA} we can write
\begin{align}
  \nonumber  &\beta_\textsc{edr}(\Omega,\sigma,\beta)
  \\ &
  =\beta+\frac{\|\chi'\|_{_2}^2}{2\Omega}\left(1+2\frac{\partial_\Omega\ft W(\Omega,\beta)}{\beta\ft W(\Omega,\beta)}\right)\left(\frac{\beta}{\sigma}\right)^2
    +\mathcal{O}\left(\frac1{\Omega^4\sigma^4}\right),
\end{align}
which expanded to next to leading order in $\beta\Omega$, acquires the  form
\begin{align}\label{pipopotamo}
  \nonumber  &\beta_\textsc{edr}(\Omega,\sigma,\beta)\\
  &=
 \beta\bigg[1-\frac{\|\chi'\|_{_2}^2}{12}\bigg(1-\frac{(\beta\Omega)^2}{60}+\mathcal O\big(\beta^4\Omega^4\big)\bigg) \left(\frac{\beta}{\sigma}\right)^2\bigg]\nonumber\\&+\mathcal{O}\left(\frac1{\Omega^4\sigma^4}\right).
\end{align}
From this expression, we see that the dependence with the detector gap $\Omega$ appears only in higher orders. This means that to a fairly good approximation, the EDR remains exponential also for finite times, although with a modified temperature $\beta_\textsc{edr}(\Omega,\sigma,\beta)$ that depends nontrivially on the KMS temperature and on the interaction time, but is mostly independent of the gap frequency.

We will see in the following sections---where we will not rely on the series expansion \eqref{pipopotamo}---that the same weak dependence of $\beta_\textsc{edr}$ on the detector's gap $\Omega$ is present without restricting the switching function and also in the short time regimes.

\section{The Anti-Unruh effect and the KMS condition}
\label{piposec:au-kms}

We will now analyze the phenomenology first reported in \cite{Anti-Unruh} under the name of {\it Anti-Unruh} phenomena. As we discussed in the introduction, the surprising result found in \cite{Anti-Unruh} was that for short times \textit{``the transition probability  of  an  accelerated  detector  can  actually
decrease
with acceleration''.} Besides finding these results, the authors of~\cite{Anti-Unruh} further discussed that if one were to define an EDR effective temperature, $\beta_{\textsc{edr}}$, as in \eqref{otrapipocosa}, one would find that $\beta_{\textsc{edr}}$ appears to be almost independent of the detector gap, a characteristic feature of the scenarios where the KMS condition is satisfied for  detectors interacting for long times (see \eqref{piporemark}).

To shed some light on this Anti-Unruh effect, let us introduce the following two definitions:

{\bf Weak Anti-Unruh:} We define the weak Anti-Unruh regime as the set of values of the detector gap $\Omega$, interaction times $\sigma$, and KMS temperatures $T_{\textsc{kms}}=1/\beta$ for which the detector's response function decreases as the KMS temperature increases, i.e.
\begin{equation}\label{pipochorrada}
 \partial_\beta\mathcal F(\Omega,\sigma,\beta)>0.
\end{equation}
In plain words: the detector detects fewer field excitations when the temperature increases.

{\bf Strong Anti-Unruh:} We define the strong Anti-Unruh regime as the set of values  of the detector gap $\Omega$, interaction times $\sigma$, and KMS temperatures  $T_{\textsc{kms}}=1/\beta$ for which the  effective EDR temperature $T_{\textsc{edr}}$---defined in \eqref{otrapipocosa}---decreases as the  KMS temperature increases, i.e.
\begin{equation}\label{strongpipochorrada}
 \partial_\beta\beta_{\textsc{edr}}<0.
\end{equation}

Notice that the latter definition becomes important if  $\beta_\textsc{edr}$, which is defined through the detector's EDR, does not depend very strongly on $\Omega$. This was the case in \cite{Anti-Unruh}. In fact, as we discussed above and as we will see later in this section, this is very generally the case.

We will see below that it is possible to have weak Anti-Unruh phenomena and yet not observe strong Anti-Unruh phenomena, but is the opposite possible? Let us analyze more closely the relationship between the two: 
\begin{equation}
  \partial_\beta\beta_{\textsc{edr}}<0\quad\Leftrightarrow \quad\frac{\partial_\beta\mathcal F(\Omega,\sigma,\beta)}{\mathcal F(\Omega,\sigma,\beta)}>\frac{\partial_\beta\mathcal F(-\Omega,\sigma,\beta)}{\mathcal F(-\Omega,\sigma,\beta)}.
\end{equation}
From this condition we can directly see that strong Anti-Unruh implies weak Anti-Unruh, except if the following two conditions are simultaneously satisfied:
\begin{equation}\label{pipondicion}
\partial_\beta\mathcal F(-\Omega,\sigma,\beta)<0,\qquad  \partial_\beta\mathcal F(\Omega,\sigma,\beta)<0.
\end{equation} 

Since $\ft C(\omega,\beta)$ is an odd function of $\omega$ as discussed above, from \eqref{eq:FOmega} and \eqref{eq:wcp} we can see that
\begin{align}\label{pipomputator}
\partial_\beta&\mathcal F(\pm\Omega,\sigma,\beta)\!=\!-\frac1{2\pi}\!\int \!\d\bar\omega |\ft \chi(\bar\omega)|^2\Big( \mathcal{P}\left(\pm\Omega\pm \tfrac{\bar\omega}{\sigma},\beta\right)\\
&\times\partial_\beta\ft C(\Omega+ \tfrac{\bar\omega}{\sigma},\beta) +\ft C(\Omega+ \tfrac{\bar\omega}{\sigma},\beta) {\partial_\beta\mathcal{P}\left(\pm\Omega\pm \tfrac{\bar\omega}{\sigma},\beta\right)} \Big).\nonumber
 \end{align}
Since $\text{sgn}\big(\partial_\beta \mathcal{P}(\omega,\beta)\big)=-\text{sgn}\big(\partial_\beta \mathcal{P}(-\omega,\beta)\big)$, \eqref{pipomputator} shows that, if the commutator is independent of the KMS parameter, then
\begin{equation}
 \text{sgn}\big(\partial_\beta\mathcal{F}(\Omega,\sigma,\beta)\big)=-\text{sgn}\big( \partial_\beta \mathcal{F}(-\Omega,\sigma,\beta)\big),
\end{equation}
and \eqref{pipondicion} cannot be fulfilled. In other words, if the detector moves along a  trajectory that does not depend on the KMS temperature, the absence of weak Anti-Unruh also implies the absence of strong Anti-Unruh.

 In view of \eqref{eq:FOmega}, a necessary condition for the weak Anti-Unruh condition \eqref{pipochorrada} to hold is that the Fourier transform of the Wightman function $\ft W$ has to grow as $\beta$ increases somewhere in its domain.

As we discussed above, if the KMS condition is satisfied, $\ft W$ is the product  \eqref{eq:wcp} of the Planckian distribution and the Fourier transform of the commutator. Therefore, the necessary condition for having weak Anti-Unruh phenomena when the KMS condition is satisfied can be simply written as 
\begin{equation}\label{Piperrak}
 \partial_{\beta}\big(\ft C(\omega,\beta)\mathcal P(\omega,\beta)\big)<0.
\end{equation}

Let us analyze the relationship between the KMS condition and the presence of Anti-Unruh phenomena in two different scenarios: \textit{a)} when the commutator is independent of the KMS parameter (e.g., inertial detectors in a thermal background), and \textit{b)} when the commutator depends explicitly on the KMS parameter (e.g., accelerated detectors coupled to the vacuum of a massive field).

\subsection{$\ft C$ does not depend on the KMS parameter}

If $\ft C$ does not depend on the KMS parameter, we see from  \eqref{Piperrak} that the necessary condition for weak Anti-Unruh phenomena is
\begin{equation}
 \ft C(\omega)\partial_{\beta}\mathcal P(\omega,\beta)<0.
\end{equation}
From \eqref{Piploanckiana}, we know that
\begin{equation}
\text{sgn}\big(\partial_{\beta}\mathcal P(\omega,\beta)\big)=-\text{sgn}(\omega).
\end{equation}
This implies that the  necessary condition for weak Anti-Unruh  \eqref{Piperrak}  can be simplified  in this case as
\begin{equation}\label{Pipotada}
 \omega \ft C(\omega)>0.
\end{equation}
From \eqref{pipotador} we see that \mbox{$\ft C(-\omega)= -\ft C (\omega)$}, which means that $\omega\ft C(\omega)$ is even.  On the other hand, since $\ft W$ is positive~\cite{fewsterjorma}, from \eqref{eq:wcp} we see that $\text{sgn}\,\ft C(\omega)=-\text{sgn}\,\omega$. Therefore   $\omega \ft C(\omega)<0$ for all $\omega\in \mathbb{R}$ and the condition \eqref{Pipotada} will never be satisfied.

This leads to the following general result: For KMS states with respect to a timelike vector $\partial_\tau$ generating trajectories for which the commutator is independent of the KMS temperature there is no weak Anti-Unruh effect. In other words, the probability of detector excitation is monotonically increasing with the KMS temperature. This is the case of the following examples:
\begin{itemize}

\item An inertial detector coupled to a thermal state  of a scalar field of mass $m\ge 0$  in arbitrary spatial dimensions, even in the presence of an IR cutoff $\Lambda$.  Explicitly, in this case the commutator is 
\begin{equation}
 \ft C(\omega)\!=\!-
\frac{\pi^{\frac{2-d}{2}}\text{sgn}(\omega)}{2^{d-1}\Gamma(d/2)}(\omega^2-m^2)^{^{\frac{d-2}{2}}}\!\Theta(|\omega|-m)\Theta(|\omega|-\Lambda),
\end{equation}
where we recall $m$ is the field mass, $d$ is the number of spatial dimensions and $\Lambda$ is an IR cutoff.

Note that this is also true for a scalar field in a finite volume imposing Dirichlet or periodic boundary conditions (the latter as long as we do not neglect the contributions coming form the zero mode~\cite{PhysRevD.90.024015}).

\item Uniformly accelerated detectors coupled to the vacuum state of a massless scalar field in $d=1$ or $d=3$ spatial dimensions.  In these cases it can be shown \cite{Takagi} that for $m=0$, $\Lambda=0$, the commutator is the same as in the inertial case thus leading to the same conclusion. The critical relationship between an IR cutoff or a field mass and the Anti-Unruh phenomena will be further analyzed below.

\end{itemize}

Since in these cases the commutator is independent of the KMS temperature, the fact that there is no weak Anti-Unruh effect implies that there is no strong Anti-Unruh effect either. In other words, the EDR temperature increases monotonically with the KMS temperature.

\subsection{$\ft C$ depends on the KMS parameter}

Although the commutator is independent of the field state, it may still depend on the KMS temperature through the trajectory $\mathsf{x}(\tau)$. In this case it is not straightforward to derive a general result as in the previous case. Let us consider some critical examples. If the field state is the Minkowski vacuum of a scalar field, trajectories with constant acceleration $a\ge 0$, yield Wightman functions that satisfy the KMS condition with KMS temperature $a/(2\pi)$. 

For the massive case, however, the commutator depends explicitly on the acceleration \cite{Takagi}. Indeed, the Wightman function has a nontrivial dependence on \mbox{$\beta=2\pi/a$}: 
\begin{equation}
\ft W_d (\omega,\beta)=\frac{\beta e^{-\frac{\beta  \omega }{2}}}{2\pi^2  } \!\int\!\frac{\d^{d-1}\bm k}{(2\pi)^{d-1}}\left|K_{\ii \frac{\beta\omega}{2\pi}}\left(\frac{\beta}{2\pi}\sqrt{m^2+\bm k^2}\right)\right|^2,
\label{pipobesselgen}
\end{equation}
which, for $d\geq 2$  becomes
\begin{align}
\ft W_d (\omega,\beta)=&\frac{\beta e^{-\frac{\beta  \omega }{2}}}{2^{d-1} \pi^{\frac{d+3}{2}} \Gamma(\frac{d-1}{2}) }\\
\nonumber&\;\;\;\times\int\d|\bm k|\, |\bm k|^{d-2}\left|K_{\ii \frac{\beta\omega}{2\pi}}\left(\frac{\beta}{2\pi}\sqrt{m^2+\bm k^2}\right)\right|^2\!,
\label{pipobesselgen}
\end{align}
while for $d=1$ the expression \eqref{pipobesselgen} reduces to
\begin{equation}
\ft W_1 (\omega,\beta)=\frac{\beta e^{-\frac{\beta  \omega }{2}}}{2\pi^2  }  \left|K_{\ii \frac{\beta\omega}{2\pi}}\left(\frac{\beta m}{2\pi}\right)\right|^2.
\label{pipo2D}
\end{equation}

In these cases the necessary condition \eqref{Piperrak} for weak Anti-Unruh can be fulfilled. In fact it is easy to check explicitly that this condition can actually be satisfied both in the $1+1$D and $3+1$D cases. Let us first focus on the 1+1D case.

\begin{figure} 
\centering
\includegraphics[width=0.45\textwidth]{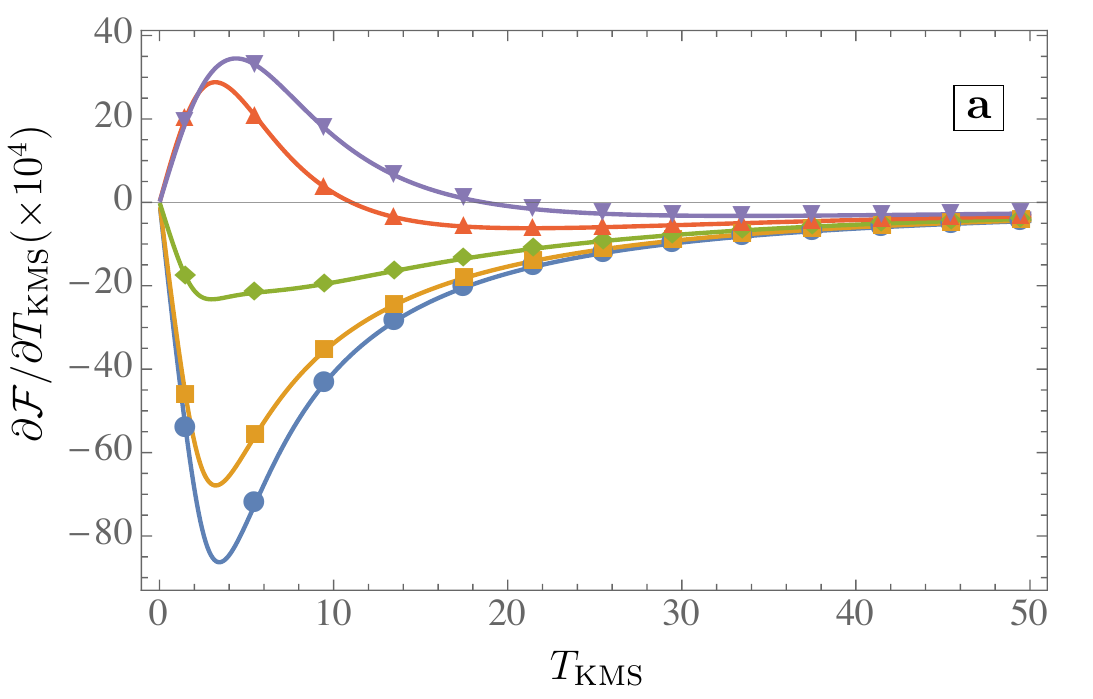}
\includegraphics[width=0.45\textwidth]{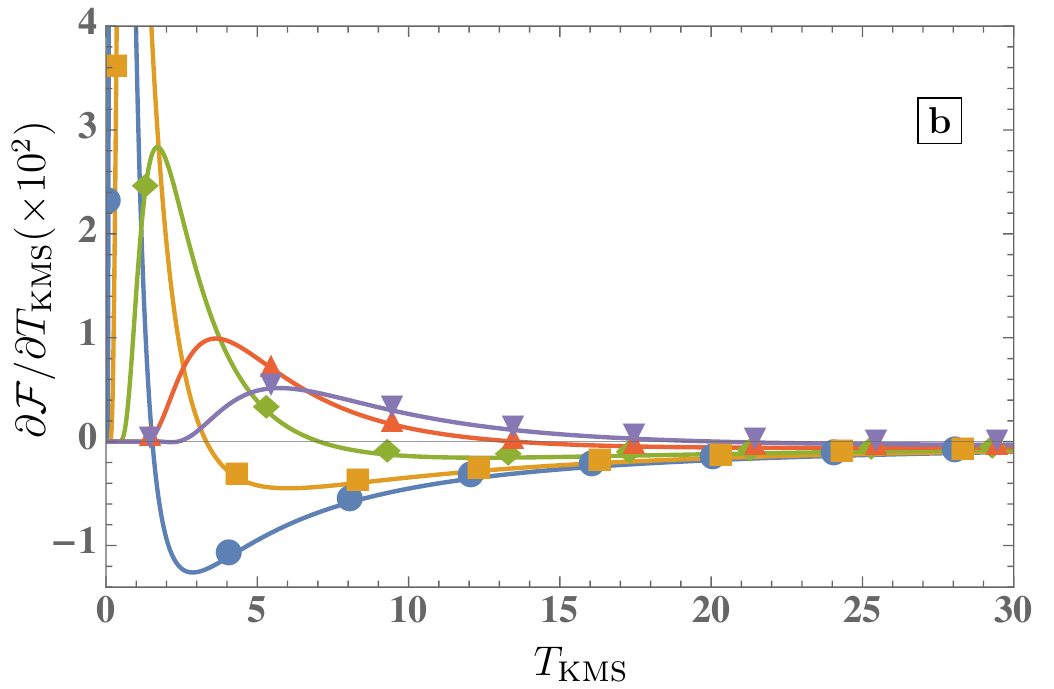}
\caption{\textit{Weak Anti-Unruh effect}: Derivative of the response function with respect to the KMS temperature $T_\textsc{kms}=1/\beta$ for 1+1D, $m=1$. The different lines correspond to values of $\Omega=15$ (inverted purple triangles), $\Omega=10$ (red triangles), $\Omega=5$ (green rhombi), $\Omega=2$ (orange squares), $\Omega=0.5$ (blue circles). The two plots represent the short and the long time regimes. Namely {\bf Top.} $\sigma=1$ with a Gaussian switching $\chi(\tau/\sigma)=\pi^{-1/4}e^{-\tau^2/(2\sigma^2)}$, and {\bf Bottom.} $\sigma\rightarrow\infty$, independently of the switching (this can be analytically evaluated through \eqref{hastalospipuevos}). We see that for a broad range of the parameters this derivative is negative (i.e., the response function decreases as the KMS temperature increases), even for adiabatic (eternal) switching. 
}
\label{fig:scheme}
\end{figure}

In the massive 1+1D case we can see (along the same lines as in the massless 1+1D case with an IR cutoff studied in \cite{Anti-Unruh}) that the accelerated detector experiences the weak Anti-Unruh effect: That is, the detector's response function can decrease as the KMS temperature $T_\textsc{kms}=1/\beta$ increases, as illustrated in Fig.~\ref{fig:scheme}a for a Gaussian switching  $\chi(\tau/\sigma)=\pi^{-1/4}e^{-\tau^2/(2\sigma^2)}$, and in Fig.~\ref{fig:scheme}b for any square integrable switching $\chi(\tau/\sigma)$ in the infinitely adiabatic limit $\sigma\rightarrow\infty$. 

Remarkably, this voids one of the major possible criticisms that could have been raised against the relevance of the Anti-Unruh phenomena reported in \cite{Anti-Unruh}. Namely, it could have been argued that in \cite{Anti-Unruh}, the introduction of a hard IR cutoff, which, rigorously speaking, yields non-stationary Wightman functions, was the responsible for the appearance of transients that give rise to the Anti-Unruh effect. However, we see  that we do not require a breakdown of the KMS condition to see the Anti-Unruh effect. Specifically, an accelerated detector coupled to a massive field vacuum will experience the weak Anti-Unruh effect in spite of the fact that the KMS condition is satisfied in this case. In other words, we can have a  detector that, when switched on for finite times, can decrease its transition rate as the KMS temperature increases.

More so, this weak Anti-Unruh behaviour also shows up even in the limit of detectors adiabatically switched on for an infinite amount of time.  Indeed, in this limit, we know that the expression of the response function is particularly simple, as shown in~\eqref{hastalospipuevos}. We show in Fig. \ref{fig:scheme}b  that the weak Anti-Unruh effect is present in the strict limit $\sigma\to\infty$, independently of the particular form of the switching function $\chi$ (even including non-adiabatic switchings for which the transition rate is well defined). Therefore the weak Anti-Unruh effect cannot be associated to transient behaviour.

\begin{figure} 
\centering
\scalebox{1.18}{\!\!\!\includegraphics[width=0.20\textwidth]{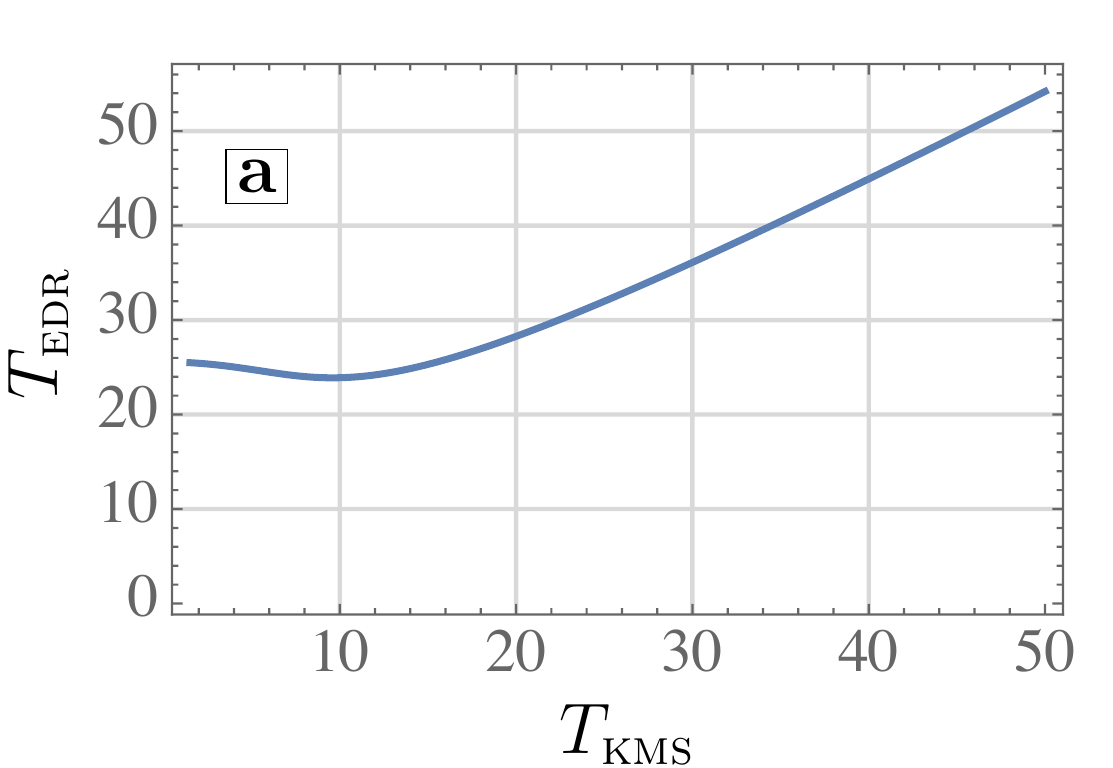}\quad
\includegraphics[width=0.197\textwidth]{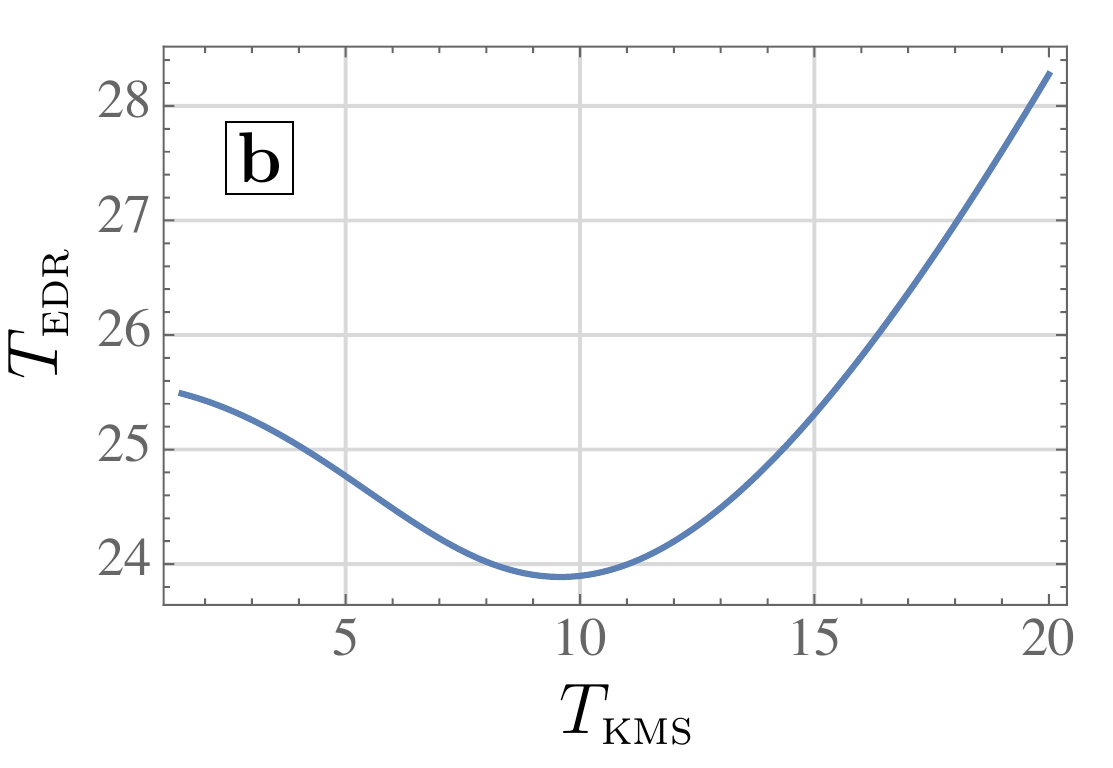}}\\[1.2mm]
\scalebox{1.17}{\!\!\!\includegraphics[width=0.20\textwidth]{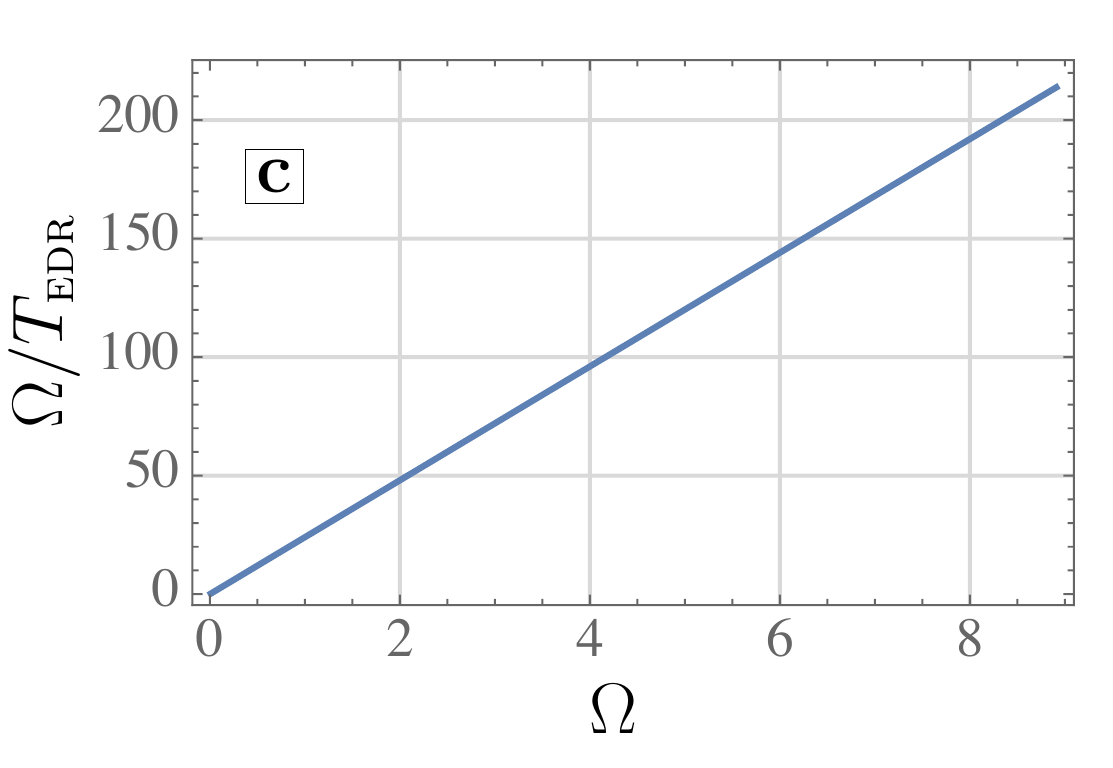}\quad \!\!\!\!\!\!\!
\includegraphics[width=0.22\textwidth]{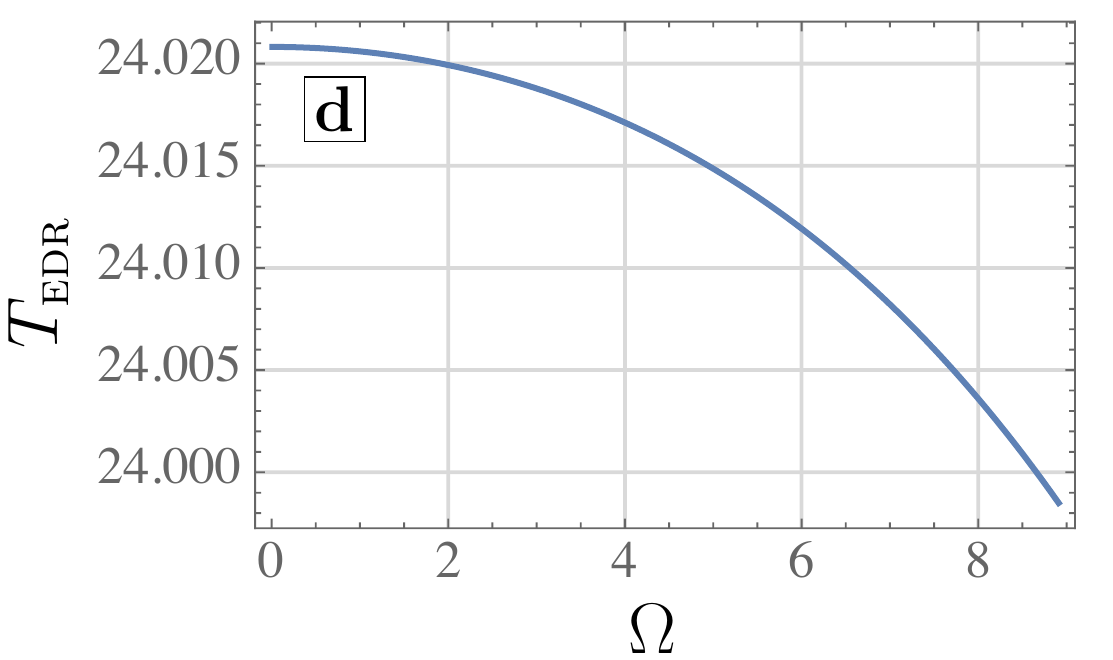}}
\caption{\textit{Strong Anti-Unruh effect}: Figure \textit{a)} shows the  EDR temperature $T_\textsc{kms}=1/\beta$ as a function of the KMS temperature for 1+1D, $m=1$, $\sigma=0.04$, and $\Omega=1$. For large KMS temperatures, $T_\textsc{edr}\simeq T_\textsc{kms}$, while for small ones the EDR temperature actually decreases, as seen in the zoomed Fig.~\textit{b)}. Figure \textit{c)} displays (for $m=1$, $\sigma=0.04$, and $T_\textsc{kms}=8$) the apparent linearity of $\Omega/T_\textsc{edr}$ with $\Omega$ and hence the almost independence of the EDR temperature with $\Omega$. Figure \textit{d)} shows (for the same parameters as Fig.~\textit{c)}) that this dependence is actually present although it is extremely weak.}
\label{fig:pipostrong}
\end{figure}

The strong Anti-Unruh behaviour, on the other hand, is confined to short interaction times and small accelerations (i.e. KMS temperatures), as shown in Fig.~\ref{fig:pipostrong}. In the figure we see that in the regime of small KMS temperatures, the EDR temperature decreases as the KMS temperature increases. We also see that for larger KMS temperatures, the EDR temperature approaches the KMS temperature, as it should be from \eqref{piporemark}. Finally, this figure also shows that the EDR temperature depends very weakly on the gap frequency $\Omega$, despite the detector not being in  equilibrium with the field. This behaviour is entirely the same as that found in \cite{Anti-Unruh}. There, a hard-IR cutoff (either removing the zero mode in a periodic cavity, or imposing a cutoff $\Lambda$ in the continuum case) causes the Wightman function not to satisfy the KMS condition but $\beta_{\textsc{edr}}$ as defined in \eqref{otrapipocosa} behaves as a function of acceleration exactly in the same way described above.

In particular, we have proven that this is a genuine effect of the acceleration of the detector, even when KMS is satisfied, and that it would not be seen by an inertial detector interacting for a finite timescale with a thermal bath regardless of the number of spacetime dimensions and the presence of cutoffs. 

We have also discussed that in 3+1D (for accelerated detectors and massive fields)  the necessary condition for weak Anti-Unruh is satisfied. However, it is not sufficient and it is still unclear whether the system  will exhibit the Anti-Unruh effect. 

\section{Parameter space dependence of the Anti-Unruh phenomena}

In this section we analyze in more detail in what region of the parameter space we can find Anti-Unruh phenomena.

One legitimate question that one may ask is whether this effect may be related with the fact that even though the response of a static detector in a thermal bath  and the response of an accelerated detector coupled to the field vacuum are statistically identical, the two responses come from fundamentally different physical effects.

In the inertial thermal case, the main contribution to the detector's excitation rate for sufficiently long times comes from rotating-wave contributions (those involving processes where the detector gets excited by emitting a field quantum \cite{scullybook}). However, in the Unruh effect, the contribution of the rotating-wave and counter-rotating wave terms (the detector gets excited by emitting a photon \cite{scullybook}) are comparable. This is the fundamental difference in the two processes and this is ultimately the reason why the two scenarios are different despite the fact that in both cases the detectors display a thermal response.

One manifestation of this intrinsic difference is the fact that the Unruh effect can excite a detector for masses below the detector's energy gap, while a thermal bath cannot. The Anti-Unruh effect is another remarkable manifestation of this difference.

In the light of this, the question could be asked whether the relationship between the detector gap scale and the mass of the field is what rules the appearance of the \mbox{Anti-Unruh} phenomena.

To answer this question, let us first consider the response function of an accelerated detector coupled to a massive field prepared in the vacuum state in the long time limit ($\sigma\to\infty$). As we showed in  \eqref{hastalospipuevos}, the response function is given by the Wightman function evaluated at $\omega=\Omega$. Specifically, for the 1+1D case the response function  is given by \eqref{pipo2D} evaluated at $\omega=\Omega$.

Let us consider two different asymptotic limits of this equation, the large mass limit and the small mass limit. 

Let us begin with the the large mass limit. Using the leading order of the asypmtotic expansion of the Bessel function for large values of its argument
\begin{equation}
    K_{\ii \frac{\beta\Omega}{2\pi}}\left(\frac{\beta m}{2\pi}\right)\sim\frac{\pi}{\sqrt{\beta m}}e^{-\frac{\beta\Omega}{2\pi}},
\end{equation}
which is valid under the condition
\begin{equation}
      \left(\frac{\beta \Omega} {2\pi}\right)^2\!\!+\frac14 \ll  \frac{\beta m}{2\pi},
\end{equation}  
we get   the following response function in the limit $\sigma\rightarrow\infty$ \cite{Takagi}:
\begin{equation}
 \mathcal{F}(\Omega,\beta)\approx \frac{e^{-\beta(\Omega/2+m/\pi)}}{4\pi m}.
\label{responsePipar}
\end{equation}

The response function \eqref{responsePipar} is a monotonically  increasing fucntion of the temperature, thus does not exhibit any kind of Anti-Unruh phenomena. This allows us to reach to the conclusion that in the asymptotic limit of field mass much larger than the detector gap for constant KMS temperature,  there cannot be any Anti-Unruh phenomena.

\begin{figure} 
\centering
\includegraphics[width=0.45\textwidth]{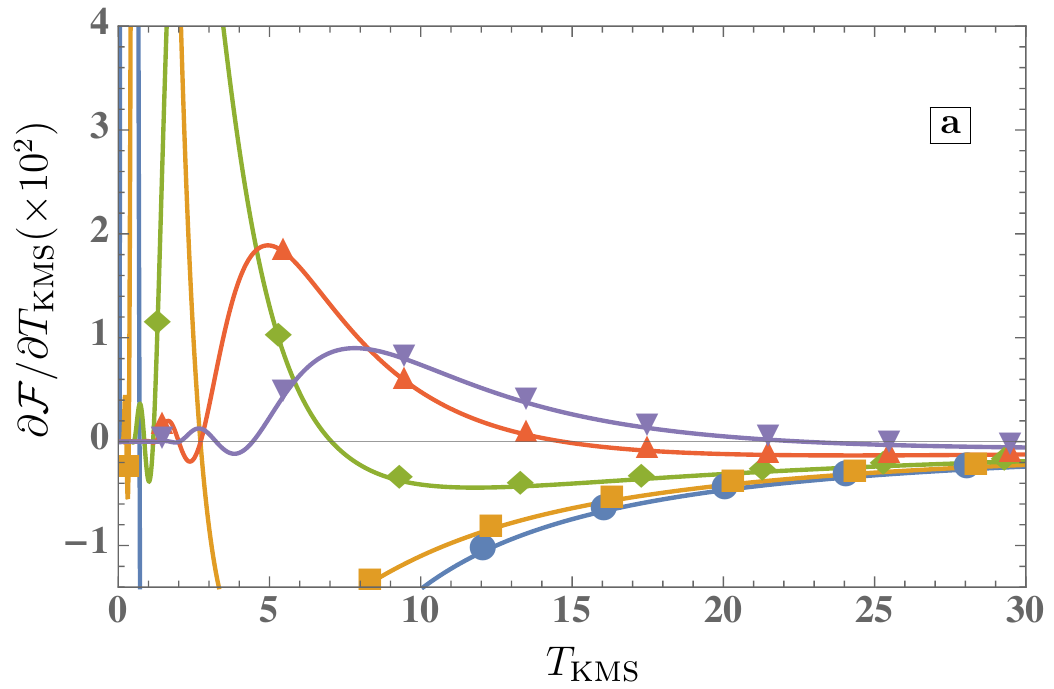}
\includegraphics[width=0.45\textwidth]{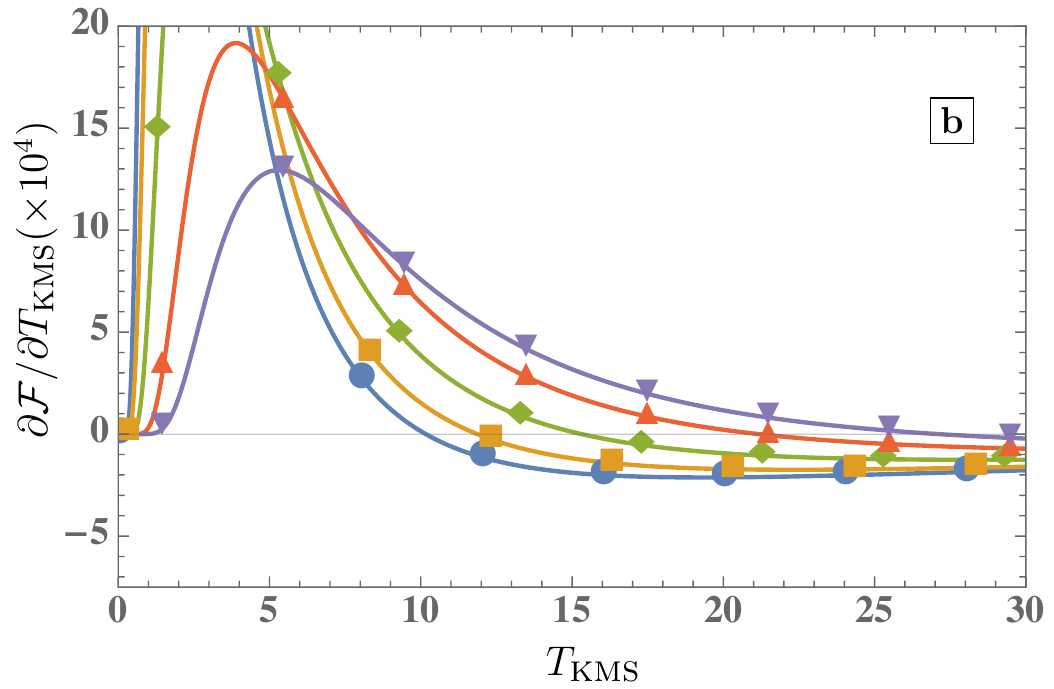}
\caption{\textit{Anti-Unruh effect dependence on the mass}: For any switching function shape, in the limit of infinite interaction time $\sigma\rightarrow\infty$, we show the derivative of the response function with respect to the KMS temperature $T_\textsc{kms}=1/\beta$ for 1+1D and for $m=0.1$ (Top) and $m=10$ (Bottom). The different lines correspond again to values of $\Omega=15$ (inverted purple triangles), $\Omega=10$ (red triangles), $\Omega=5$ (green rhombi), $\Omega=2$ (orange squares), $\Omega=0.5$ (blue circles). The top figure shows how for $ m \ll T_\textsc{kms}$ the oscillations of the derivative generate Anti-Unruh effect in the low temperature zone, whereas the lower shows that for for  $ m \gg T_\textsc{kms}$ the AntiUnruh effect disapperars. We see in both figures that the Anti-Unruh effect can exist for values of $\Omega$ below and above the mass scale $m$. 
}
\label{fig:schemepar2}
\end{figure}

On the other hand, as shown in \cite{Takagi}, the asymptotic behaviour of the response function in the limit of small mass is given by
\begin{equation}
\mathcal{F}(\Omega,\beta)\!\sim\!\frac{1}{\Omega(e^{\beta\Omega}-1)}\bigg[1+\cos\left(\frac{\beta \Omega}{\pi}\log{\frac{\beta m}{4\pi}}\!+\!\phi\left[\frac{\beta \Omega}{2\pi}\right]\right)\bigg],
\label{piponocturno}
\end{equation}
with $ \phi(z)=2\,\text{Arg}\,\Gamma(\ii z)$, in the regime where
\begin{equation}
    \left(\frac{\beta \Omega}{2\pi}\right)^2+1\gg \left(\frac{\beta m}{2\pi}\right)^4
\end{equation}
is satisfied.

In the light of \eqref{piponocturno}, we see that the response function in the limit of small $\beta m$, where $\beta\Omega$, is kept constant is not a monotonically increasing function of $\beta$. In fact, \eqref{piponocturno} becomes highly oscillarory as $\beta m$ goes to zero and, as such, its derivative with respect to the KMS temperature will take negative values. The Anti-Unruh phenomena will appear therefore for sufficiently small $\beta m$ regardless of the constant value of $\beta$ and $\Omega$.

The conclusion that we extract is that although there may be some relationship between the Anti-Unruh phenomena and the ratio between $\Omega$ and $m$, the existence of the Anti-Unruh effect is independent of the scale of $\Omega$, since, for sufficiently small mass, we can find Anti-Unruh phenomena regardless of the value detector gap. Instead the relevant figure of merit ruling the appearance of the phenomena is the ratio between the field mass and the acceleration.

We illustrate this in Fig. \ref{fig:schemepar2}, where we show that the anti Unruh phenomena for detectors interacting for long times ($\sigma\to\infty$) can exist for a diverse range of parameters. In particular, it can exist when $m$ is more than an order of magnitude larger than $\Omega$  (Fig. \ref{fig:schemepar2}a) and also when $m$ is more than an order of magnitude below $\Omega$ (Fig. \ref{fig:schemepar2}b). In both cases it can be seen that the  Anti-Unruh effect ceases to appear when $\beta m \gtrsim 1$,

\section{Conclusions}
\label{piposec:conc}

To shed light into the recently discovered Anti-Unruh phenomena \cite{Anti-Unruh}, we have analyzed the role of the interaction time in the thermalization of an Unruh-DeWitt particle detector. In particular, it is well-known that, for infinitely long interaction times and if the Wightman function satisfies the KMS condition, the Excitation-to-Deexcitation Ratio (EDR) is determined by the detailed balance condition. We have shown by explicitly writing a series expansion of the response function that, for long but finite interaction times, the EDR depends very weakly on the detector's energy gap $\Omega$.

This means that, even when the detector has not reached equilibrium, the effective EDR temperature \eqref{otrapipocosa} can exhibit an approximate independence of the gap. This independence has often lead  to inappropriately regarding it as a true temperature associated with some equilibrium state. Despite being independent of $\Omega$,  the EDR temperature is not the KMS temperature, but a nontrivial function of the interaction time and the KMS temperature itself. The independence of $T_{\textsc{edr}}$ on the detector's gap appears to hold  even in the very short interaction time regimes (where the series expansion discussed above is not valid). For instance, we have shown that this is the case for an accelerated detector coupled to the Minkowski vacuum of a massive scalar field at sufficiently low KMS temperature (i.e. acceleration) and sufficiently small interaction times.

The Anti-Unruh effect can be characterized in terms of the behaviour of the response function and the effective EDR temperature with the KMS temperature $T_\textsc{kms}$. On the one hand, we have called {\it weak} Anti-Unruh effect those situations in which the excitation probability decreases as $T_\textsc{kms}$ increases (a detector clicks less often as the KMS temperature of the field increases). On the other hand, we have called \textit{strong} Anti-Unruh effect situations where the effective EDR temperature---almost independent of the gap frequency---decreases as $T_\textsc{kms}$  increases.

We have seen that the weak and strong Anti-Unruh effects do  not appear at all under some general conditions. Namely, that the Wightman function satisfies the KMS condition with respect to translations along the proper time of a detector whose trajectory does not depend on the KMS temperature. 
In particular, this implies that the Anti-Unruh effects (both weak and strong) are absent for inertial detectors coupled to massless or massive scalar fields in a KMS state (for example, a thermal state) with or without a momentum cutoff and for any spatial dimensions. It is also absent for  accelerated detectors in the Minkowski vacuum of a massless scalar field for one and three spatial dimensions. 

The situation is entirely different for an accelerated detector   coupled to the Minkowski vacuum in two different but related cases, namely, there can be Anti-Unruh phenomena for a massive scalar field or when an IR cutoff is in operation. We showed that in these cases there appear clear signatures of both weak and strong Anti-Unruh behaviour in 1+1 spacetime dimensions. 

For the massive case, for small interaction times, and well within the regime of validity of perturbation theory, we see that \textit{i)} the response function  decreases as the acceleration (the KMS temperature) increases, and \textit{ii)} the effective EDR temperature decreases with the KMS temperature, depends also on the interaction time, but is almost independent of the gap frequency. Furthermore, for long interaction times or large KMS temperatures, the strong Anti-Unruh effect disappears but, remarkably, the weak version of it is still at work, i.e. the derivative of the response function with respect to the KMS temperature is negative. This is true even in the strict limit of infinitely long adiabatic switching for any square integrable switching function.

The massless case with an IR momentum cutoff was studied in \cite{Anti-Unruh} with the same results. It must be stressed that although in \cite{Anti-Unruh}, the Wightman function was not stationary and hence was not KMS, the Anti-Unruh effect cannot be associated with this fact because it is also present in the massive case, which certainly is KMS. This effect cannot be  dismissed as a transient either since, as we have discussed the (weak) Anti-Unruh effect (i.e., a detector `seeing less particles' as the temperature of the medium increases) is present even for infinitely long times.

Finally, let us note that the fact that the Anti-Unruh effect can be seen by relativistic accelerated observers but not by inertial observers coupled to a thermal bath is a distinctive signature of perceived particle creation by accelerated observers, that can be singled out from the behaviour of detectors coupled to thermal backgrounds.

\acknowledgments

L. J. G. was partially supported by the Spanish MINECO through the projects  FIS2014-54800-C2-2 (with FEDER contribution). E. M-M acknowledges the funding of the NSERC Discovery program. The authors thank Allison Sachs for her comments on the final version of this manuscript. The authors also thank Jorma Louko for his always insightful remarks.

\appendix

\section{Infrared cutoff and the KMS condition}

In this appendix we show that the introduction of an IR cutoff makes it impossible to satisfy the KMS condition. Let us consider  a momentum cutoff $\Lambda$. Then, the Wightman function  in the Minkowski vacuum state  for any spatial dimension $d$ has the functional form
\begin{equation}
    W_d(\mathsf y)=\int \frac{\d \mathsf k^{d+1}}{(2\pi)^{d+1}} \Theta(k^0)\delta(\mathsf k^2+m^2)\,e^{\ii \mathsf k\cdot \mathsf y}\Theta(|\bm k|-\Lambda),
\end{equation}
where  $\mathsf k\cdot \mathsf y$ is the spacetime contraction of the spacetime vectors $\mathsf k$ and $\mathsf y$. We now perform  a spatial rotation to a frame such that $y^i=0$ for all $i=2,3\ldots$ and a Lorentz boost along the direction $y^1$ to the rest frame. If $\mathsf k'$ is the momentum in this new frame, the Wightman function acquires the form 
\begin{align}
 W_d(\mathsf y)&=\int \frac{\d \mathsf k'^{d+1}}{(2\pi)^{d+1}} \Theta(k'^0)\delta(\mathsf k'^2+m^2)
 \nonumber\\
 &\times e^{-\ii k'^0\text{sgn}(y^0)\sqrt{(y^0)^2-(y^1)^2}}\Theta(|\bm k(\mathsf k',\mathsf y)|-\Lambda),
\end{align}
where
\begin{equation}
 |\bm  k(\mathsf k', \mathsf  y)|^2=\frac{(-y^1 k'^0+y^0 k'^1)^2}{(y^0)^2-(y^1)^2}+\bm k'^2- (k'^1)^2 .
 \label{pipokk}
\end{equation}
The pullback of the Wightman function along a linear trajectory with uniform acceleration is obtained by evaluating it at $\mathsf y=\Delta \mathsf x=\mathsf x(\tau)- \mathsf x(\tau')$. We see that the integrand of $W_d(\Delta \mathsf x)$ is invariant under time translations
except perhaps for the factor $(-\Delta \mathsf x^1 k'^0+\Delta \mathsf x^0 k'^1)^2$ in \eqref{pipokk}.
Indeed $\text{sgn}(\Delta x^0)$ is obviously invariant under time translations and so is the proper interval $(\Delta x^0)^2-(\Delta x^1)^2=(4/a^2)\sinh^2\big(a(\tau-\tau')/2\big)$ for accelerated 
trajectories. Finally, $(-\Delta \mathsf x^1 k'^0+\Delta \mathsf x^0 k'^1)^2$  is invariant for inertial trajectories but not for not vanishing accelerations: in the former case both $\Delta x^0$ and $\Delta x^1$ are proportional to the time difference, while in the latter they are nontrivial combinations of hyperbolic functions of the time parameter. Since the Wightman function is not time-translationally invariant, the KMS condition cannot be satisfied.

\bibliography{biblio}

\end{document}